\documentclass[twocolumn,showpacs,preprintnumbers,amsmath,amssymb]{revtex4}
\usepackage{epsfig,bm, graphicx, latexsym}
\usepackage{amsmath,epsf,psfig}

\begin{document}
\newcommand{\be}{\begin{equation}}
\newcommand{\ee}{\end{equation}}
\newcommand{\la}{\langle}
\newcommand{\ra}{\rangle}

\title{Theory of Aging in Structural Glasses}

\author{Vassiliy Lubchenko} \altaffiliation[Current Address:
]{Department of Chemistry, Massachusetts Institute of Technology,
Cambridge, MA 02139.}  \author{Peter G. Wolynes}

\affiliation{Department of Chemistry and Biochemistry and Department
of Physics, University of California, San Diego, 9500 Gilman Drive, La
Jolla, CA 92093-0371}

\date{\today}

\begin{abstract}

The random first order transition theory of the dynamics of
supercooled liquids is extended to treat aging phenomena in
nonequilibrium structural glasses. A reformulation of the idea of
``entropic droplets'' in terms of libraries of local energy landscapes
is introduced which treats in a uniform way the supercooled liquid
(reproducing earlier results) and glassy regimes. The resulting
microscopic theory of aging makes contact with the
Nayaranaswamy-Moynihan-Tool nonlinear relaxation formalism and the
Hodge-Scherer extrapolation of the Adam-Gibbs formula, but deviations
from both approaches are predicted and shown to be consistent with
experiment. The nonlinearity of glassy relaxation is shown to
quantitatively correlate with liquid fragility. The residual
nonArrhenius temperature dependence of relaxation observed in quenched
glasses is explained. The broadening of relaxation spectra in the
nonequilibrium glass with decreasing temperature is quantitatively
predicted.  The theory leads to the prediction of spatially
fluctuating fictive temperatures in the long-aged glassy state, which
have non-Gaussian statistics. This can give rise to ``ultra-slow''
relaxations in systems after deep quenches.

\end{abstract}

\maketitle

\section{Introduction}

The energy landscape metaphor has turned theorists towards viewing the
global geometry of the phase space of complex systems. When these
systems are mesoscopic in size, for example, small proteins
\cite{OLSW} or gas phase clusters \cite{Wales}, a more or less
complete mathematical formulation of the idea, capable of treating
kinetics and thermodynamics, can be made using statistical tools to
characterize minima and saddle points of the entire system. Yet for
macroscopic systems, most transitions rearrange particles only
locally.  This essential aspect of the dynamics is brought home
forcefully by noting that a liter of liquid water will move from one
energy minimum to another in $10^{-39}$ sec. Such a short time scale
cannot be directly relevant to any laboratory measurement on this
system!  The necessity for using a local description is clearly
recognized in the modern theory of supercooled liquids and glasses
which is based on the statistical mechanics of random first order
transitions \cite{KW_PRA}-\cite{LW_soft}. In the deeply supercooled
liquid regime this theory explains not only the phenomenological
features of the dynamics \cite{KTW} but quantitatively predicts, on a
microscopic basis, the size of cooperative lengths, the precise
non-Arrhenius behavior of typical relaxation times \cite{XW,LW_soft}
and the non-exponentiality of relaxation \cite{XWbeta}. A quantized
version of the theory explains the low temperature thermodynamics of
amorphous substances, usually interpreted in terms of two level
systems \cite{LW} and the more energetic Boson peak excitations
\cite{LW_BP}. The crucial manifestation of the locality concept in
this theory (which has many mean field, global, aspects) is the notion
that ``entropic droplets'' drive the large-scale activated notions in
glass forming liquids \cite{KW_PRB,KTW,XW} and give both liquids and
glasses an intrinsic ``dynamical mosaic'' structure.

The main purpose of this paper is to describe the predictions of the
RFOT theory for the behavior of glasses that have fallen out of
equilibrium because of being rapidly cooled from the melt. Dynamics
does not cease in rapidly quenched liquids that become glasses; rather
motions persist but are frustratingly slow and hard to study
experimentally. This slow, far-from-equilibrium dynamics, called
``aging'', is not only of fundamental interest for statistical
mechanics \cite{Bouchaud} but also is of great practical interest
since so many amorphous materials are used in everyday life for times
exceeding their preparation time \cite{Struik}. Sometimes even small
changes of properties upon aging are crucial to engineering
performance.

Like the quantum theory of the low temperature properties of glasses,
the description of the nonequilibrium aging regime requires the
explicit construction and study of the energy landscape of local
regions of the glass. This local energy landscape description turns
out to be a microcanonical ensemble reformulation of the entropic
droplet concept that was formulated originally in the canonical
ensemble. We have found this microcanonical description to be more
vivid and somewhat easier to communicate than the original canonical
formulation. After introducing this formulation we will show that it
indeed reproduces the results for the dynamics of equilibrated
supercooled liquids already obtained using the random first order
transition theory. More important, this local energy landscape theory
makes several very striking predictions about the aging regime and how
the aging dynamics in the nonequilibrium glass is related to the
kinetics and thermodynamics of the equilibrated liquid. The
predictions are consonant with all experiments known to us and make
contact with, but are formally distinct from, the phenomenological
approaches commonly used to describe aging in structural glasses
\cite{Tool,Narayanaswamy,Moynihan,Scherer,Hodge}. The present theory,
for example, makes a specific prediction of the so-called nonlinearity
parameter, $x_{NMT}$, in the Narayanaswamy-Moynihan-Tool, formalism
\cite{Tool,Narayanaswamy,Moynihan}. In addition, the degree of
nonexponentiality of relaxation characterized by a $\beta$ exponent is
predicted. $\beta$ turns out to be neither precisely fixed at $T_g$
nor does it precisely scale like the $\beta$ for an equilibrated
liquid with $T$, another assumption often made. The predicted
variation of $\beta$ with quench temperature is however modest until
very low temperatures are reached. Likewise the Adam-Gibbs expression
\cite{AdamGibbs} has been extended phenomenologically notably by Hodge
and Scherer into the aging regime below $T_g$ by assuming the
configurational entropy to be frozen at $T_g$. While above $T_g$, an
Adam-Gibbs like form of the temperature dependence of relaxation time
is found from the present theory, below $T_g$, the local energy
landscape approach gives a different expression for the relaxation
rate in an aging glass. Quantitatively this expression gives
relaxation rates close to the Hodge-Scherer-Adam-Gibbs (HSAG) latter
extrapolation but the local energy landscape theory predicts a
deviation from that formula. This deviation would be interpreted
within the HSAG framework as a quench temperature dependence of the
apparent configurational entropy. Such a deviation has been found by
Alegria et al. in their careful work on aging of polymers
\cite{Alegria}.

Our most explicit results are obtained for the idealized situation of
the aging initially found after a cooling history with a single rapid
quench of modest depth. We will also qualitatively discuss
modifications of the simple theory expected for very deep quenches.
We also discuss the behavior of systems that have significantly
relaxed in the quenched state. While, our analysis suggests that, to a
first approximation, introducing a single fictive temperature should
serve well to describe many quench histories, the local energy
landscape theory shows that using a single fictive temperature is not
exact. We suggest possible modifications of the usual aging kinematics
based on the present theory.

The organization of the paper is as follows: we first describe how the
local energy landscape view of entropic droplets can be visualized and
show how a library of local energy landscapes can be constructed. We
next show how to derive in this framework the (previously obtained)
Vogel-Fulcher behavior above $T_g$. We then present the results for
typical relaxation rate for rapidly quenched, aging glasses and
compare these predictions with experimental results. We then discuss
the predictions for the stretching exponent both above and below
$T_g$. Following this, the generalizations needed for very deeply
quenched glasses and glasses that have undergone significant
relaxation in the aging state are discussed. Finally we summarize our
theory and suggest some further experimental tests of it.

\section{The Local Energy Landscape Construction}

The energy landscape language is usually applied to a small system
(protein or cluster). It is also used for a thermodynamically large
system described by mean field theory \cite{MPV,Coluzzi}. In the first
situation, the barriers between states are, of course, finite because
the system is finite, but barriers may be formally infinite for the
mean field system. The ``states'' for a cluster or a protein are often
taken to be the basins surrounding minima of the potential energy
\cite{inherent}. These are well defined and no transitions can occur,
classically, between them at absolute zero. The ``states'' of a mean
field system are tied to minima of free energy and again, owing to the
$O(N)$ barriers, no transition can occur between them (even at finite
$T$).

In a supercooled liquid the observed plateau in the time dependent
neutron scattering correlation function \cite{Mezei} shows that most
molecules spend a great deal of time vibrating about a given
location. At least deep in the supercooled regime (where the plateau
is well developed) we can, therefore, conceptually imagine
constructing an average location for any particle about which it
vibrates (for times less than the plateau). The three dimensional
structure based on these average locations will be quite close to a
potential energy minimum found by removing kinetic energy from the
system with infinite speed (``steepest descent'' to an ``inherent
structure''). A mean field theory of the glass transition can be
obtained by constructing a free energy functional dependent on the
(plateau time averaged) density and then noting that this functional
has minima for density patterns localized around such inherent
structures \cite{SW,SSW,DasguptaValls}. In this way a vibrational
component of the free energy can be defined and can be added to the
average energy of such a state to give a free energy which determines
the thermal probability of being in this state through the Boltzmann
law.

Free energy functionals only have such aperiodic minima below a
temperature $T_A$. This temperature has been shown (within a simple
approximation) to be equivalent to the mode coupling dynamical
transition temperature \cite{KW_PRA}. The transition has the character
of a spinodal for a first order phase transition \cite{KW_PRB}. Above
$T_A$ even simple vibrational motions will take the system from one
minimum to another, but below $T_A$, the plateau in the time dependent
structure factor indicates the persistence of such states and
therefore their relevance to dynamics. In fact, as a first
approximation, the persistence of these states allows them to act as a
``basis set'' for describing supercooled liquid dynamics. Thus we say
a free energy landscape emerges at $T_A$. $T_A$ has been evaluated
both for simple liquids \cite{SW,SSW} and for models of network
forming liquids with repulsive force \cite{HW} from first principles.
The physical meaning of the temperature $T_A$ is simple. When
supecooled below its melting temperature $T_m$, the liquid is, of
course, in a metastable state as a whole, and therefore must be
metastable {\em locally}. $T_A$ can be conveniently interpreted, for
example, as a temperature below which every two molecules will spend a
particular number (say, 300) of vibrational periods together between
the first encounter and the final departure.

As part of our programme to describe the microscopics of liquid
relaxations, we will show that transitions between the metastable
configurations of the liquid {\em as a whole} consists of transitions
between {\em local} metastable configurations. We will employ a
microcanonical procedure in which only local regions are considered
and will find that beyond a certain (relatively small) size $N^*$, the
thermal and relaxational properties of the liquid do not depend on the
size of the sample. Alternatively said, all liquid properties of
interest can be deduced by focusing on a liquid region of size $N^*$
and completely disregarding what the rest of the liquid is doing.
This is the essence of the {\em locality} of the liquid free energy
landscape. The microcanonical procedure is a necessary step in
establishing such locality; it is however rarely used, so let us first
train our intuition on the very familiar example of a harmonic
lattice. Imagine being inside an extremely large, cubic, and purely
harmonic lattice and being given the ability to do arbitrary
thermodynamic measurement locally.  You are further assigned to
explain, within a formal model, those thermal measurements, but are
allowed to visually inspect and mechanically test the {\em bonds} with
an arbitrarily large, but {\em finite} region, limited by how fast you
can perform the inspection. Upon checking that all individual bonds
are truly harmonic and the lattice is cubic within a certain region
(of size $L$), you write down a simple hamiltonian but are left with
the issue of the conditions at the boundary - strictly speaking, the
assignement is undoable!  Being an optimist, you say: let me {\em
assume} for today that the rest of the sample is totally rigid (fixed
boundary conditions) or does not exist at all (open bondary condition)
and diagonalize the resulting Hamiltonian; tomorrow I can expand my
horizon, repeat the procedure and see what happens. The first day
proves frustrating though, because the Hamiltonian has an energy gap
(proportional to $\pi/L$), while the thermal measurements clearly show
very low frequency excitations are present. Fortunately, since
(unbeknownst to you) the lattice was harmonic and periodic, your
consecutive inspections and diagonalizations will yield a smaller and
a smaller gap. Clearly, the landscape of a regular harmonic lattice is
non-local and the microcanonical procedure offers a definitive test of
non-locality. Consider yet another elementary notion: Suppose you have
a product of two Hilbert spaces $A \otimes B$ (possibly interacting
via $V_{A,B}$) with an energy $\epsilon_{i_A,j_B}$ assigned to
configurations $i_A$ and $j_B$ of the subspaces $A$ and $B$. Consider
the partition function $\sum_{i_A,j_B} e^{-\beta \epsilon_{i_A,j_B}} =
\sum_{i_A} e^{-\phi_{i_A}}$, where $e^{-\beta \phi_{i_A}} \equiv
\sum_{j_B} e^{-\beta \epsilon_{i_A,j_B}}$. If $A$ and $B$ are
independent: $V_{A,B}= 0 \Rightarrow \epsilon_{i_A,j_B} =
\epsilon_{i_A} + \epsilon_{j_B}$, - one may further simplify
$\phi_{i_A} = \epsilon_{i_A,j_B} - T S_B$. In either case,
$\phi_{i_A}$'s are {\em free} energies of the degree of freedom $B$,
but can still be (with advantage) regarded as an energy, as far as the
$A$ degree of freedom is concerned. For example, one is allowed to
build a {\em microcanonical} construction using $\phi_{i_A}$'s as the
label.  In the following, we will use a $\phi_{i_A}$-like quantity to
describe {\em all} the degree of freedom in the supercooled (or
quenched) liquid in excess of the lowest energy crystalline structure
corresponding to this substance (at this temperature). (For a polymer
that does not form a crystal our results will still apply
qualitatively.) While the ``integrated out'' degrees of freedom $B$
are clearly related to the vibrations of the corresponding crystal
structure, they are easy to conceptualize only in the mean field
limit, where they are indeed harmonic vibrations (see
shortly). Otherwise, the ``$B$'' motions are strongly anharmonic.

\begin{figure}[tb]
 \includegraphics[width=.95\columnwidth]{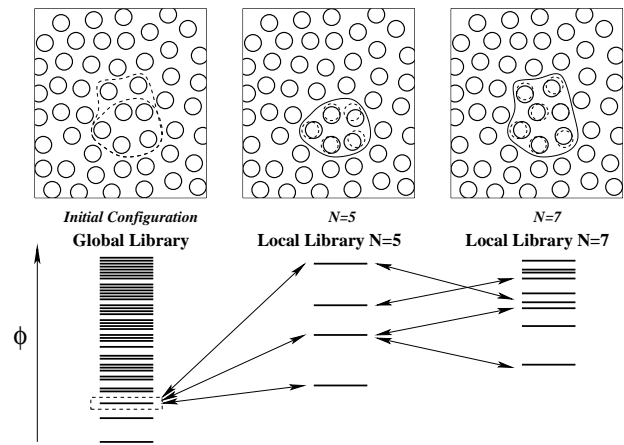}
 \caption{\label{library} In the upper panel on the left a global
configuration is shown, chosen out of a global energy landscape. A
region of $N=5$ particles in this configuration is rearranged in the
center illustration. The original particle positions are indicated
with dashed lines. A larger rearranged region involving $N=7$
particles is connected dynamically to these states and is shown on the
right. In the lower panel, the left most figure shows the huge density
of states that is possible initially. The density of states found in
the local library originating from a given initial state with 5
particles being allowed to move locally is shown in the second
diagram. These energies are generally higher than the original state
owing to the mismatch at the borders. The larger density of states
where 7 particles are allowed to move is shown in the right most part
of this panel. As the library grows in size, the states as a whole are
still found at higher energies but the width of the distribution
grows.  Eventually with growing $N$, a state within thermal reach of
the initial state will be found.  At this value of $N^*$ we expect a
region to be able to equilibrate.}
\end{figure}

We are now ready construct a library of free energy minima for a very
large sample of a liquid (or glass). This spectrum is shown
schematically in the first column of Figure \ref{library} as a
function of the free energy of a state $i$, $\phi_i^{lib}$ which we
take as the sum of an energy $\epsilon_i$ and an entropic contribution
from vibrations within the basin $- T s_{vib,i}$.

Because the system is very large this global library of states has an
exceedingly dense spectrum, whose density dramatically increases with
free energy. Let us imagine the system is presently in one of these
basins with (free) energy $\phi_i^{lib}$. We can now construct a set
of {\em local} libraries of states. To do this, imagine mentally
cutting out a region around a location ${\bf R}$ containing $N$
particles, where $N$ is much less than the total number of particles
in the sample. Call this region $D({\bf R}, N)$. Next, freeze the
molecules outside this region but allow molecules within $D$ to
move. (Unlike in the cubic crystal analogy, a supecooled structure has
built-in stress, so using open boundary condition is not useful.)
With the frozen environment one could (by quenching the potential
energy or a free energy functional) find a new set of inherent
structures that involve only reconfiguring the particles in
$D$. Actually such a set of structures (apart from strains near the
surface of $D$) will locally resemble a subset of the original global
library. If $N$ is small, however, we expect the subset of the states
so sampled to be selected on average from a higher energy part of the
global spectrum than where the original $\phi_i^{lib}$ was
found. Essentially this is because the density of states rapidly
increases with $\phi$ and there is less freedom to readjust particles
in the constrained minimization problem because of the fixed
environment. In absolute terms, we expect the discrepancy in energy of
the mean energy of the states in any local library from the starting
state to increase with $N$ since it is the particles at the borders
that must be most strained.

When $N$ is small the local library is sparse but it grows denser and
spreads out over a larger range of free energy as the size of the
region increases. The set of such libraries centered around ${\bf R}$
is shown also in Figure \ref{library} for $N=1$ up to a large value of
$N$. The number of states in the library at free energy $\phi_i$
determines the configurational entropy $\Omega_c(\phi_i^{lib}) =
e^{S_c(\phi_i^{lib})/k_B}$, where $k_B$ is the Boltzmann constant. The
competition between the average energy growth and the spreading of the
range of energies with size of the region means that there is a
characteristic size $N^*$ where a state will finally be found within
thermal energy of the starting state. $N^*$ turns out to be the size
of a dynamically correlated region in the liquid or glass. Since the
new configuration is within $k_B T$ of the starting state, a
transition of such a region from its original configuration to the new
state can occur with reasonable probability. The region after
reconfiguration will be characterized by a temperature $T$. Elementary
transitions in the liquid must leave most of the molecules near their
old locations - this is the dynamical essence of locality. Therefore
the local landscape libraries are locally connected to each other: in
order to re-arrange a large region, smaller regions located in the
same place must first be rearranged. The specific law connecting
states in neighboring libraries must obey detailed balance (at the
vibrational temperature $T$) but otherwise will depend specifically on
molecular details for each system. The additional activation energy,
for a downhill move in such a library should only be a few times the
energy needed to heat a particle to $T_A$, the dynamical transition
temperature.

Irrespective of the detailed motions allowed, the locality of the
dynamics, however, guarantees that the transitions from the initial
state to one of the thermally allowed states in the library for
$D({\bf R}, N^*)$ will nevertheless be slow: owing to the initial rise
of the average energy there will be a bottleneck in the probability
flux at an intermediate value of $N$, namely $N^\ddagger$. The
activation free energy for reaching this bottleneck state determines
the escape rate from the initial configuration by motions in the
vicinity of ${\bf R}$.

While the local libraries can be constructed explicitly given
sufficient computational resources, the RFOT theory suggests a useful
set of approximations to the statistics of these libraries when $N^*$
and $N^\ddagger$ are moderately large. We now describe these
approximations: We ascribe to each state $j$ of the library for a
region $D({\bf R}, N)$ a so-called bulk free energy
$\Phi_j^{bulk}({\bf R}, N)$. Naively this would be the sum of the
vibrational entropies and internal energies of each molecule in $D$
and the pair interactions between the molecules within $D$. (Rigor
here would require ensuring the appropriate finite ranges of
interactions and some precisely specified ways of partitioning
vibrational entropies: entanglement entropies of the interface are too
subtle for consideration at present). Note that in a system with a
globally correlated landscape, such as the harmonic lattice considered
earlier, a description in terms of local libraries will not provide
the full list of configurations available to the system as a whole.

In a similar way we would define a bulk free energy of the initial
state relevant to this region $\Phi_{in}^{bulk}({\bf R}, N)$. The
actual energy of the complete sample when the state $j$ is inserted in
region $D({\bf R}, N)$ will not just reflect the difference in the
bulk energies corresponding to the state $j$ and the initial state;
instead, it will be higher by an amount $\Gamma_{j,in}$:
\begin{equation} 
\phi_j^{lib} - \phi_{in}^{lib} = \Phi_j^{bulk}({\bf R}, N) -
\Phi_{in}^{bulk}({\bf R}, N) + \Gamma_{j,in}.
\label{phi_j}
\end{equation}

As mentioned earlier, the new local structure labelled by
$\phi_j^{lib}$, is likely to be higher in energy than the initial
configuration. Our construction thus suggests $\Gamma_{j,in}$ will
usually be positive and will at most scale with the interface area. We
will write $\Gamma_{j,in}=\gamma_{j,in} N^x$. As we shall see later
the estimate for the exponent $x=2/3$ in 3 dimensions, based solely on
the interface area, is probably naive near $T_g$.

We now wish to calculate the equilibration rate of the region $D({\bf
R}, N^*)$. If the environment remains frozen, this rate is the escape
rate from $N=0$ to $N=N^*$. The probability flux to increase $N$ falls
until the bottleneck value $N^\ddagger$ is reached. After
$N^\ddagger$, even though the average energy of states in the
libraries increases, the growth of the number of states with
increasing library size is sufficient so that a rapid path to a
thermally equilibrated state at size $N^*$ can be found. Let the
typical downhill microscopic rearrangment have a rate
$\tau^{-1}_{micro}$. This rate will only be weakly activated. The flux
to any state at a size $N$ smaller than $N^*$ will be
\begin{eqnarray} 
k & = & \tau^{-1}_{micro} \int (d \phi_j^{lib}/c_\phi)
e^{S_c(\Phi_j^{bulk})/k_B} e^{-(\phi_j^{lib}-\phi_{in}^{lib})/k_B T}
\nonumber \\ & \simeq & \tau^{-1}_{micro}
e^{S_c(\Phi^{bulk}_{eq})/k_B} e^{-(\phi_{eq}-\phi_{in}^{lib})/k_B T},
\label{k}
\end{eqnarray}
where $\phi_{eq}$ maximizes the integrand and $c_\phi$ is some
constant of units energy. The quantities $\phi_j^{lib}$ and
$\Phi_j^{bulk}$ are related through Eq.(\ref{phi_j}). This
maximization means $\phi_{eq}$ will be the internal free energy
characteristic of the system at the ambient (i.e. vibrational)
temperature $T$. That is, $\phi_{eq}$ is the sum of the energetic and
vibrational entropic contributions appropriate to equilibrium at
$T$. In essence, the equation above is nothing more than the rate of
escape through a transition state with non-zero entropy, hence a
structure resembling the canonical ensemble. Aside from a numerical
factor, the integration in Eq.(\ref{k}) is indeed a canonical
summation in terms of $\phi_j$, which should be regarded as the
non-meanfield analog of the free energy of the so called ``pure''
state.  The concept of the pure state is well developed in the context
of frustrated mean-field spin systems \cite{MPV}. In the mean-field
limit, the pure states are separated by infinite barriers and thus the
vibrations around the metastable free energy minima are purely
harmonic.  The quantity $\phi_j^{lib}$ is non-meanfield because it can
be defined only locally (after paying the ``price'' of the surface
energy $\Gamma_{j,in}$).  It is the degrees of freedom due to
transitions between those (non-meanfield) ``pure'' states of the
liquid that give rise to the (measurable) configurational
entropy. Note also that owing to the intense and strongly anharmonic
motions in the liquid (at this finite $T$), a local liquid state
labelled by a particular value of $\phi$ is not a single inherent
structure \cite{inherent}, but rather is a weighted superposition of
many inherent structures, with more or less harmonic vibrations on
top.

\begin{figure}[tb]
\includegraphics[width=.55\columnwidth]{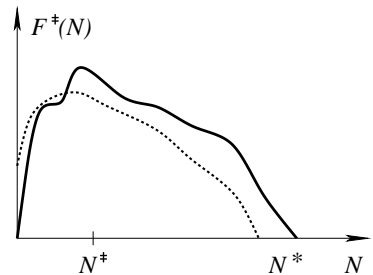}
\caption{\label{profiles} The free energy to reconfigure the initial
configuration is shown as a function of the size of the locally
rearranged region. There will be fluctuations in the shape owing to
the detailed packing found in the initial configuration but on the
average the profile is given by equation (\ref{F(N)fluc}) in the
text. The dashed curve shows the profile for an initial state which is
much higher than the equilibrium energy at $T$, while the solid curve
is the free energy profile for an initially equilibrated state.  }
\end{figure}

Note, $\phi_{eq}$ is a function of $T$ and $N$. Since
$\Phi^{bulk}_{eq}$ is the equilibrium bulk free energy at temperature
$T$, one may replace $S_c(\Phi^{bulk}_{eq})$ by its equilibrium value
at that temperature $S_c(N,T)$. Thus we get for the escape rate to $N$
the result
\begin{equation}
k(N) = \tau^{-1}_{micro} \exp\left\{S_c(N,T) -
\frac{\phi_{eq}-\phi_{in}^{lib}}{k_B T} \right\}.
\end{equation}
The location of the bottleneck is determined by the minimum of this
expression over $N$.  We can thus define an activation free energy
\begin{equation}
F^\ddagger(N) = \phi_{eq} - \phi_{in}^{lib} - T S_c(N,T),
\label{F_phi}
\end{equation}
whose maximum defines the bottleneck location. Notice this activation
barrier depends on the total free energy of having any target state at
size $N$ and the initial nonequilibrated particular state free energy
$\phi_{in}^{lib}$ in which we only include a vibrational contribution
and no configurational entropy. This function is shown in
Fig.\ref{profiles}.
Introducing Eq.(\ref{phi_j}) into Eq.(\ref{F_phi})
yields:
\begin{equation}
F^\ddagger(N) = F_{eq}^{bulk}(N,T) -\Phi_{in}^{bulk}(N,T) + \gamma
N^x,
\label{F(N)fluc}
\end{equation}
where $F_{eq}^{bulk}(N,T) = \Phi^{bulk}_{eq} - TS_c(N,T)$ is the
total equilibrium free energy that includes both the configurational
entropy of the region of size $N$ and the internal bulk free energy
with its vibrational contribution.  In this expression we have
substituted the average mismatch energy coefficient $\gamma$, for the
state specific values.  For dynamically connected states
$F_{eq}^{bulk}(N,T)$, $\Phi_{in}^{bulk}(N,T)$, and $S_c(N,T)$ grow
linearly with $N$.  There will also be $\delta F$, an additional
fluctuation that typically scales as $N^{1/2}$. Thus we write
\begin{equation} 
F^\ddagger(N) = [f_{eq}(T) - \phi_{in}(T)] N + \gamma N^x + \delta F,
\end{equation}
where $f_{eq}(T) = F_{eq}^{bulk}(N,T)/N$ is the total bulk free energy
per particle of the final state at temperature $T$ and $\phi_{in}(T)=
\Phi_{in}^{bulk}(N,T)/N$ is the internal free energy per particle of
the initial state. In the following, we will omit the fluctuation term
$\delta F$, whenever computing the most probable barrier, but will
consider it explicitly when estimating the degree of
non-exponentiality of the relaxation, which is directly related to the
barrier fluctuations. Note, the expression above is simply what would
be prescribed by a regular nucleation theory for the free energy
barrier of conversion from a (usually non-equilibrium) initial state
to the other (usually equilibrium) state. Clearly, when the initial
state is at equilibrium: $\Phi^{bulk}_{in} = \Phi^{bulk}$, - the
``free energy'' driving force $f_{eq}(T) - \phi_{in}(T)$ is equal to
$-T s_c$, i.e. there is still relaxation in the supercooled liquid, as
driven by conversion between alternative (local) aperiodic packings of
the liquid. In general, the maximum of the typical $F^\ddagger$ occurs
at $N^\ddagger$ such that $\partial F^\ddagger/\partial
N|_{N=N^\ddagger}$ vanishes giving
\begin{equation} 
N^\ddagger =
\left(\frac{\phi_{in}-f_{eq}}{x\gamma}\right)^{\frac{1}{x-1}}
\label{N(x)}
\end{equation}
and a typical (most probable) rate
\begin{equation} 
k = \tau_{micro}^{-1}
\exp\left\{-\frac{\gamma}{T}\left(\frac{\phi_{in}-f_{eq}}{x\gamma}
\right)^{\frac{x}{x-1}} (1-x) \right\}.
\label{k(x)}
\end{equation}

We finish this Section by elaborating on why the relaxations occuring
with the rate from Eq.(\ref{k(x)}) can occur with a zero free energy
gap thus leading to a local free energy landscape. Consider first a
supercooled liquid at equilibrium (i.e. not quenched):
$f_{eq}-\phi_{in} = -Ts_c$. Clearly the size $N^*$ at which
$F^\ddagger (N^*)=0$ corresponds to the same liquid state, therefore a
region of size $N^*$ can survey {\em all} liquid configurations
typical of this temperature. An isoenergetic state exists because the
system resides in a metastable state much higher than the lowest
energy crystalline arrangement, as reflected in non-zero
configurational entropy $s_c$, and the very high density of states
$e^{S_c(\phi)}$. (Note, this notion underlies the existence of
residual structural degrees of freedom in glasses even at cryogenic
temperatures \cite{LW}.) If the liquid is quenched and aging, the
driving force $f_{eq}-\phi_{in}$ is still negative (it will be
computed shortly), so both quantities $N^\ddagger$ and $N^*$ exist and
are comparable to their equilibrium values, although $N^*$ no longer
signifies straightforwardly the degree of the landscape locality
(consistent with the system being out of equilibrium in the first
place). Regardless, relaxations are local during aging too.

\section{Relaxation in the Equilibrated Supercooled Liquid}

Examining the expressions for the relaxation rate for an equilibrium
sample makes explicit the connection to the earlier form of the
entropic droplet idea. In the equilibrated sample case, the internal
vibrational free energies of the initial and final states are the same
so the driving force to distinct equilibrated configurations comes
only from the configurational entropy contribution available at the
equilibrium temperature. People are sometimes confused how
configurational entropy, which is never made available in a strict
mean field scenario, can drive a transition event. This is because of
the locality of the landscape in the droplet analysis, which goes
beyond meanfield thinking. Locally, for a region of space $D({\bf R},
N^*)$ there is a ``funnel'' of states leading away from the initial
configuration to other equally equilibrated
configurations. Configurational entropy drives the activated motions
of the local regions of a glass in the same way the entropy arising
from the large number of denatured configurations of a protein drives
the unfolding of a folded native protein even though the denatured
states are individually higher in energy \cite{OLSW}. Any individual
escape path is likely to find a big barrier but this barrier is
partially cancelled by the growth of the number of escape routes. The
free energies along each path fluctuate thus increasing the likelihood
of finding a low energy path, when a large number of paths is
available. As the entropy per particle, $s_c(T)$ gets smaller the
driving force to re-equilibrate falls so the critically activated
region of the glass necessary for re-equilibration grows larger and
the rate falls, owing to the larger activation barrier from the
mismatch contribution. The precise way this happens depends on the
mismatch energy and its exponent $x$.

Therefore, before proceeding to study the nonequilibrium situation, we
digress to discuss aspects of the average mismatch energy $\Gamma$
which we have approximated as $\gamma N^x$. First it is clear that the
form of $\gamma N^x$ is only a crude approximation to the mismatch
energy when $N$ is small. In principle, recall, the mismatch energy
could be explicitly computed by carrying out the local library
construction on a computer. The only problem is obtaining initial
configurations equilibrated to the appropriate low
temperatures. Finding such equilibrated configurations deep in the
landscape currently requires heroic computer resources. Therefore the
mismatch energy must presently be inferred by analytical
considerations.

The simplest free energy functional calculations of the mismatch
energy gives an energy proportional to the interface area and
therefore gives the mismatch exponent $x=2/3$.  Such simplified
calculations also give an explicit value of the prefactor of the
scaling relation, $\gamma_0$ at the ideal glass transition
temperature, $T_K$. At this temperature, the lost interaction terms of
a sharp interface have to balance the entropy cost of localizing each
particle to its cage.

This gives, at $T_K$ \cite{XW,LW},
\begin{equation}
\gamma_0 = \frac{2\sqrt{3\pi}}{2} k_B T_K
\ln\left[\frac{(a/d_L)^2}{\pi e}\right],
\end{equation}
where $d_L$ is the mean square fluctuations of particles in a given
basin \cite{inherent} and $a$ is the interparticle spacing. The ratio
$d_L/a$ is about 0.1 for glassy configurations, just as it is in the
Lindemann criterion for melting. Calculations based on free energy
functional show that $\gamma$ vanishes as $T_A$ is approached from
below \cite{KW_PRB}. This is because $T_A$ resembles a spinodal. From
these estimates, $\gamma(T)$ can be obtained \cite{LW_soft}. These
naive mean field estimates would give an activation barrier varying as
$s_c^{-2}$, as first detailed by Kirkpatrick and Wolynes \cite{KW_PRB}
and later discussed by Parisi \cite{Parisi}.

Kirkpatrick, Thirumalai and Wolynes pointed out an effect left out in
the naive estimate of the mismatch energy \cite{KTW}: One must
acknowledge there are numerous solutions of the mean field equations
describing minima of the free energy functional, these precisely
correspond to our local energy landscape libraries. In principle some
of these other configurations can be interpolated between the internal
target state to which the region is relaxing and its fixed environment
in order to lower the mismatch energy. This interpolation is called
``wetting''.

Such wetting only can be rigorously defined for very large
$N$. Wetting is a dynamical process that takes some time to
develop. We should therefore for greater accuracy ascribe a frequency
dependence to the mismatch surface tension $\gamma(T,\omega)$. The
mapping between the free energy functional and the random field Ising
model allows us to invoke an argument of Villain \cite{Villain} that
gives a curvature dependence to the surface tension coefficient
$\gamma=\gamma_0 (a/R)^{1/2}$ in three dimensions where $a$ is the
microscopic length where the surface tension $\gamma_0$ is established
($R^d \propto N$). Accounting for this wetting correction leads to the
exponent $x=1/2$. When this exponent is used in Eq.(\ref{k(x)}), we
see that the relaxation rate in the equilibrium system has exactly the
Adam-Gibbs form
\begin{equation}
k=\tau_{micro}^{-1} e^{-A/s_c},
\label{k_AG}
\end{equation}
where $s_c$ is the configurational entropy per particle. If we assume
that $s_c$ vanishes linearly at an ideal glass transition temperature
$T_K$, this rate agrees with the Vogel-Fulcher law. Distinct from the
AG argument, however, in the RFOT theory the critical size
$N^\ddagger$ scales as $s_c^{-2}$ not $s_c^{-1}$ and the dynamical
correlation length is much bigger than the AG picture implies, but if
the simple estimate of $\gamma_0$ based on the vibrational free energy
cost is used, explicit values of the typical barrier height are
obtained in addition to the experimental scaling of rates, as found by
the RFOT theory \cite{XW}. The surface tension $\gamma_0/T_g$, being a
logarithmic function of the vibrational amplitude in the glass,
depends little on the atomic make up of the glass. Therefore if we use
the Vogel-Fulcher expression
\begin{equation} 
k=k_0 \exp\{-D T_g/(T-T_K)\},
\end{equation}
the D is predicted to depend inversely on $\Delta c_p$ - the change in
heat capacity upon vitrification. Specifically one finds
\begin{equation} 
D=32k_B/\Delta c_p
\end{equation}
The resulting correlation of barrier heights (measured by $D$) and
glass thermodynamics is excellent \cite{XW}. The softening of $\gamma$
near $T_A$ also explains very well the deviations from the VTF laws
that are observed as the temperature is raised to $T_A$ where there is
a crossover to collisional dynamics \cite{LW_soft}. We will use the
same ``wetted'' form of the surface energy term without softening in
the body of the paper to follow. We must note however even at $T_g$,
these microscopic calculations give $N^*$ only of the order of
hundreds of particles (consistent with experiment \cite{Ediger}), so
we are far from asymptopia and other ``ultimate'' scalings (closer to
$T_K$) are conceivable.

The local library and mismatch energy concepts let us discuss {\it en
passant} various ``defect'' pictures of glassy dynamics in landscape
terms. Many theories of the glassy state imagine there is a basic
undefected structure at the heart of the phenomenon (an
``ur-structure'').  Doubtless for many systems, the periodic crystal
itself is one such basic ``ur-structure''. For big enough $N^*$ the
periodic crystal will indeed be a member of the local landscape
library but it will not be entropically favored and as long as the
surface tension between liquid and crystal remains sufficient it will
not be a major target state. Certainly ``devitrification'' can and
does occur in the laboratory but we will leave the study of this
transformation for future discussion. Other ur-structures have been
discussed as dominating glassy dynamics such as icosahedral crystals
\cite{Nelson} or other ``avoided crystalline phases''
\cite{Kivelson}. Deep in the local energy landscape, these structures
and their defected forms must be found. The dimension of the defects
supported by these ur-structures will determine the mismatch energy
and thus how the rate will depend on the driving entropy, in the
present picture. Point defects such as interstitial-vacancy pairs have
an energy cost independent of the region site i.e. giving an exponent
$x=0$ and would give a relaxation rate independent of
$s_c$. NonArrhenius behavior based on point defects usually must rely
on special kinetic constraints, which must be encoded in the
transition rules \cite{Andersen}. By construction, the
quasi-equilibrium estimate of the rate made above would fail for such
models. Of course if there were only a few ``dead-end'' states these
could be explicitly subtracted out in the estimates. Such a situation
may apply for entangled polymers.

Simple estimates of the free energies of such point defects puts their
energy cost near to the limit seen in ordinary laboratory glass
transitions \cite{Eastwood}. At low temperature they may conceivably
short-circuit the generic transitions proposed here. Line-like defects
enter prominently into the constellation of approaches based on
frustrated icosahedral order \cite{Nelson} and presumably should occur
also when other types of frustrated regular ordered systems act as
ur-structures. Nussinov has argued that hamiltonians for uniformly
frustrated systems should exhibit random first order transitions in
the mean field approximation \cite{Nussinov}. It is likely correct to
view the states accessible in the high density of states region
relevant for real glasses in this way even for models based on
frustrated order.

Recall, in assessing the relevance of defect based models, that the
measured configurational entropy per particle at the laboratory $T_g$
is about 1 $k_B$. This means we are usually far from the low defect
density regime.

At low defect density, nevertheless, line-like defects may either
traverse the correlated region directly or wander across the region
like a Brownian path. The line length in the first case of
``ballistic'' traversal should scale like $N^{1/3}$ while in the
Brownian case the length is proportional to square of the traversed
distance so we expect the mismatch energy to scale like $N^{2/3}$,
resembling the mean field result without wetting. It is interesting
that an interpolation between the ballistic and Brownian scalings
would be hard to distinguish from the result we use $\Gamma \sim
N^{1/2}$. Even at $T_g$, $N^*$ is only about 150 so all these power
laws (with the exception of the point defect case) could likely be
fitted to a detailed mismatch construction with similar accuracy for
liquids in the laboratory.

\section{Relaxation in the Immediately Quenched Glass}

The microcanonical local landscape arguments allow us to estimate the
relaxation rates once the statistics of the energies of the local
regions in the initial nonequilibrium state are known. Once the system
has fallen out of equilibrium, in general, these statistics depend on
the detailed quenching history of the sample. The complete aging
theory should determine these statistics self-consistently. In the
present and following sections we will assume the quench involves
simple straightforward cooling and that little time has elapsed since
the glass transition was passed. In this case the statistics of the
initial energies are taken to be those of an equilibrium system at a
temperature $T_g$. Presumably $T_g$ will be temperature of the
midpoint of the dynamic heat capacity drop upon cooling.  Thus we take
$\phi_{in}(T)$ to be the bulk energy at $T_g$ augmented by the
vibrational part of the free energy at the ambient temperature
$T$. The fluctuations of the internal energy of regions, which
determine the nonexponentiality of relaxation in the glass will also
be taken to be the same as the equilibrium fluctuations at $T_g$, and
therefore to be determined by $\Delta c_p(T_g)$.

Using the mismatch exponent $x=1/2$ in equations (\ref{N(x)}) and
(\ref{k(x)}) we obtained fairly simple results for the critical
cluster size
\begin{equation} 
N^\ddagger = \left[\frac{2(\phi_g-f_{eq})}{x\gamma}\right]^{-2}
\end{equation}
and the typical rate
\begin{equation} 
k = \tau_{micro}^{-1} \exp\left\{-\frac{\gamma^2}{4 k_B T
[\phi_g-f_{eq}]} \right\}.
\label{k(x)g}
\end{equation}

The initial and final states are at the same vibrational temperature
but not at the same configurational temperature. The initial state is
typical of a $T_g$ configuration but the activated state is
``equilibrated'' to the temperature $T$ both in vibrational terms and
in locally configurational terms. Because of this the driving force
for reconfiguration is not purely entropic in the nonequilibrated case
Thus there is a big conceptual difference from the usual extrapolation
of the Adam Gibbs formula. In addition, the rates predicted by
Eq.(\ref{k(x)g}) differ from the Adam-Gibbs formula extrapolated with
a fixed configurational entropy $s_c(T_g)$.

At $T_g$, clearly the rates predicted by the equilibrated formula
(\ref{k_AG}) and Eq.(\ref{k(x)g}) are the same. Upon further cooling
the nonequilibrium rate while decreasing, however, is substantially
higher than it would be at equilibrium. This is because the driving
force for reconfiguration not only includes the configurational
entropy at the ambient temperature but also an energy increment of the
initial state $\Delta \epsilon = \epsilon_g - \epsilon_{eq}(T)$. (We
assume vibrational free energy contributions are nearly the same in
inherent structures typical of $T_g$ and the ambient temperature $T$.)
Just such a change of slope is observed in the laboratory: the
apparent activation energy in the nonequilibrium glassy state is
smaller than the extrapolated value from the equilibrated supercooled
liquid. This change is usually quantified in the
Nayaranaswany-Moynihan-Tool framework through the expression
\begin{equation}
k_{n.e.} = k_0 \exp\left\{-x_{NMT}\frac{\Delta E^*}{k_B T} -
(1-x_{NMT}) \frac{\Delta E^*}{k_B T_f} \right\}. \hspace{4mm}
\end{equation}
where $E^*$ is the equilibrated apparent activation energy at $T_g$
and $x_{NMT}$ lies between 0 and 1. To compare with our initial quench
result we would take $T_f = T_g$, again noting in a realistic cooling
history $T_f$ would need to be self-consistently determined. We note
that equation (\ref{k(x)g}) gives a gradual transition to Arrhenius
behavior. We can find $x_{NMT}$ most easily at low temperatures (near
$T_K$), where the typical relaxation rate will follow an Arrhenius law
according to Eq.(\ref{k(x)g}). The Arrhenius behavior applies because,
for ordinary liquids near to $T_K$ we expect the mismatch free energy
to largely be energetic so we take it as a constant. By the definition
of the Kauzmann temperature $f_{eq}(T_K)= \phi_K$ is also mostly
energetic since the configurational entropy vanishes. The vibrational
components of $f_{eq}$ and $\phi_g$ are assumed to cancel so the
activation energy is in the nonequilibrium very low temperature regime
\begin{equation} 
\Delta E_{n.e.L.T.}^\ddagger = \frac{\gamma(T_K)^2}{4
(\epsilon_g-\epsilon_K)}
\label{EneLT}
\end{equation}
where $\epsilon_g$ is the energy per particle of the frozen glassy
state, as prepared, and $\epsilon_K$ is the energy per particle in the
putative ideal glassy ground state equilibrated at the Kauzmann
temperature. The energy difference $\epsilon_g-\epsilon_K$ is
determined by the configurational part of the heat capacity $\Delta
c_p(T)$ for intermediate values of $T$ such that $T_g > T > T_K$. The
apparent activation energy in the nonequilibrated glass at low
temperatures turns out to be comparable to the equilibrium activation
{\em free} energy, $\Delta F^\ddagger$ at $T_g$. It therefore should
not vary much from substance to substance but depends on the quenching
time scale $t_Q$ through the relation
\begin{equation} 
t_Q = \tau_{micro} \exp\left(\frac{\Delta F^\ddagger_g}{k_B T }
\right)
\label{tQ}
\end{equation} 
To obtain this result of the near equality of $\Delta F_g^\ddagger$
and $\Delta E_{n.e.L.T.}^\ddagger$ let us take the configurational
heat capacity to have the form $\Delta c_p= \Delta c_p(T_g)T_g/T$, as
suggested by Angell \cite{Angell_cp1,Angell_cp2}. This form is based
on good laboratory estimates. We can now find by integrating and
insert this into Eq.(\ref{EneLT}) to give:
\begin{equation} 
\Delta E_{n.e.L.T.}^\ddagger = \frac{\gamma(T_K)^2}{4 \Delta c_p(T_g)
k_B T_g \ln(T_g/T_K)}.
\end{equation}
The activation free energy at $T_g$, on the other hand is obtained by
finding $s_c$ from the integration of $\Delta c_p/T$ and using this in
the equilibrium rate expression to give:
\begin{equation} 
\Delta F_g^\ddagger = \frac{\gamma(T_g)^2}{4 \Delta c_p(T_g) k_B T_g
(T_g/T_K-1)}.
\end{equation}
The ratio of the activation energy in the glass to the equilibrated
free energy barrier at $T_g$ is therefore
\begin{equation} 
\frac{\Delta E_{n.e.L.T.}^\ddagger}{\Delta F_g^\ddagger} =
\left[\frac{\gamma(T_K)}{\gamma(T_g)} \right]^2
\frac{(T_g/T_K-1)}{\ln(T_g/T_K)}.
\label{DDratio}
\end{equation}
If $T_K$ and $T_g$ are close, as they are for ``fragile'' systems
expanding the logarithm gives a ratio close to one. Even for the
strong liquid Si0$_2$ the ratio of $T_g$ (1480K) to $T_K$ (876K) is
only 1.7, which gives $\Delta E_{n.e.L.T.}^\ddagger/\Delta
F_g^\ddagger=0.8$, if we neglect any temperature dependence of
$\gamma$.  For a laboratory glass transition on the one hour time
scale we should find universally to a good approximation (1 hour
quench) ~ 32 to 39 $k_B T_g$. To compare with the NTM phenomenology we
note that the RFOT theory predicts universally that the activation
energy (not free energy!!!) in the glassy state depends on the quench
time
\begin{equation} 
\frac{\Delta E_{n.e.L.T.}^\ddagger}{k_B T_g} = \alpha
\ln(t_Q/\tau_{micro})
\end{equation} 
with a coefficient $\alpha$ very close to one. This is consonant with
many experiments.

\begin{figure}[tb]
 \includegraphics[width=.6\columnwidth]{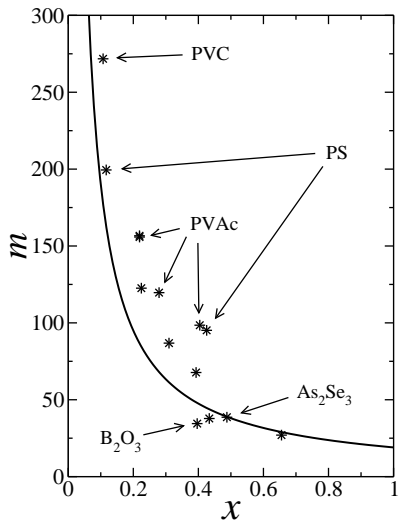}
 \caption{\label{m_x_labelled1} The fragility parameter $m$ is plotted
as a function of the NTM nonlinearity parameter $x_{NMT}$. The curve
is predicted by the RFOT theory when the temperature variation of
$\gamma_0$ is neglected. The data are taken from
Ref. \cite{app_phys_rev}. A few substances (PVAc = polyvinyl acetate,
PVC = polyvinylchloride, PS = polystyrene, B$_2$O$_3$, and
As$_2$Se$_3$) are labeled. Notice some measured values are not
consistent on multiple measurements; this may reflect a breakdown of
phenomenology for the history dependence discussed in the text or
different material preparation. The more fragile substances lie above
the prediction without barrier softening, which has no adjustable
parameters.  }
\end{figure}

The apparent equilibrium activation energy at $T_g$ in the liquid is
much larger than the nonequilibrium value found in the glassy state
(being in a sense cancelled in the rate expression by a large positive
entropy of activation). Again assuming $\Delta c_p$ depends on
temperature, we find
\begin{eqnarray} 
\Delta E_{g,app}^\ddagger &=& \left. \frac{\partial (\Delta
F^\ddagger/T)}{\partial (1/T)}\right|_{T_g} \nonumber \\ &=& \Delta
F_g^\ddagger \left\{2 \frac{\partial \ln \gamma(T)}{\partial(-\ln T)}
+ 2 + \frac{1}{T_g/T_K-1}\right\} \hspace{5mm}.
\end{eqnarray}
Using the condition that the configurational entropy vanishes at
$T_K$, this energy can also be expressed in terms of the ratio of
$c_p(T_g)$ to $s_c(T_g$):
\begin{equation} 
\Delta E_{g,app}^\ddagger = \Delta F_g^\ddagger \left\{2
\frac{\partial \ln \gamma(T)}{\partial(-\ln T)} + 2 + \frac{\Delta
c_p(T_g)}{s_c(T_g)} \right\}.
\label{DEapp}
\end{equation}
We then obtain for the inverse of the nonlinearity parameter $x_{NMT}$ at
low temperatures from Eqs.(\ref{DDratio}) and (\ref{DEapp}):
\begin{widetext}
\begin{equation} 
x_{MNT}^{-1}  = \frac{E_{g,app}^\ddagger}{\Delta E_{n.e.L.T.}^\ddagger}
= \left\{2 \frac{\partial \ln \gamma(T)}{\partial(-\ln T)} + 2 +
\frac{\Delta c_p(T_g)}{s_c(T_g)} \right\}
\left[\frac{\gamma(T_g)}{\gamma(T_K)} \right]^2
\frac{\ln(T_g/T_K)}{(T_g/T_K-1)}.
\end{equation}
\end{widetext}
$\Delta c_p(T_g)/s_c(T_g)$ is a thermodynamic measure of the liquid
fragility. This ratio is large for very fragile liquids and small for
``strong'' liquids. Thus we see ``fragile'' liquids are very nonlinear
while ``strong'' liquids, in general, should not be. Just such a
correlation has been discussed by McKenna and Angell \cite{McKenna}
and can be expressed as a relation between the kinetic fragility
parameter
\begin{equation} 
m = \left. \frac{1}{T_g} \frac{\partial \log_{10}
\tau}{\partial(1/T)}\right|_{T_g} = \frac{\Delta
E_{g,app}^\ddagger}{k_B T_g}\log_{10}e .
\end{equation}
and the nonlinearity parameter $x_{NMT}$. Using the previously derived
relation (\ref{DDratio}) and the equation above we find
\begin{eqnarray}
x_{MNT}^{-1} &=& \frac{E_{g,app}^\ddagger}{\Delta
E_{n.e.L.T.}^\ddagger} \nonumber \\ = &m& \left\{ (\log_{10}e)
\frac{\Delta F_g^\ddagger}{k_B T_g}
\left[\frac{\gamma(T_K)}{\gamma(T_g)} \right]^2
\frac{(T_g/T_K-1)}{\ln(T_g/T_K)} \right\}^{-1} \hspace{5mm}
\end{eqnarray}
Assuming the $\gamma$ ratio is near one and using the glass transition
temperature appropriate to 1 hr, so $\Delta F_g^\ddagger/k_B T_g = \ln
10^{17} \simeq 39$, and a generic $T_g/T_K = 1.26$ \cite{LW_soft}, we
obtain
\begin{equation}
m \simeq \frac{19}{x}.
\end{equation}
This relation is plotted in Figure \ref{m_x_labelled1} along with
data for several systems. We see that the estimate agrees reasonably
well with experiment. In this estimate we have neglected the $T$
dependence of $\gamma$. The microscopic treatment from RFOT shows
however that $\gamma_0$ depends both on the proximity to $T_K$ and to
$T_A$, as discussed in our paper on the barrier oftening effect
\cite{LW_soft}. Thus we will generally have both an entropic and an
energetic contribution to $\gamma_0$ which may explain some of the
scatter in the curves. Indeed we see more fragile systems lie
systematically above the curve as is expected since $T_K$ and $T_A$
are closer leading to the longer barrier softening effect.

According to the RFOT theory the typical relaxation time in the
nonequilibrium quenched state does {\em not} immediately become
Arrhenius in temperature dependence below the laboratory glass
transition. Thus there should be deviations from the NMT formalism,
which might be crudely fit by allowing the nonlinearity parameter to
be $T$ dependent. Indeed some of the scatter in
Fig.\ref{m_x_labelled1} probably arises also from this cause.  Below
$T_g$, the expected nonArrhenius behavior from RFOT theory is much
weaker than the divergently nonArrhenius behavior found above
$T_g$. Within the Adam-Gibbs nonequilibrium extrapolation advocated by
Hodge \cite{Hodge} and Scherer \cite{Scherer}, the RFOT theory result
would appear to involve a temperature dependent ``configurational
entropy''. Just such a behavior has been found by Alegria et al. in
polymer systems \cite{Alegria}. One way to express this connection is
to compare $\Delta F_{n.e.}^\ddagger(T)$ to its value at $T_g$. The
inverse of this ratio would be the apparent configuration entropy in a
Hodge and Scherer-Adam-Gibbs extrapolation. In view of
Eq.(\ref{k(x)g}), the inverse ratio is given by:
\begin{eqnarray} 
\left[\frac{\Delta F_{n.e.}^\ddagger(T)}{\Delta
F_g^\ddagger}\right]^{-1} = \left[\frac{\phi_g-f_{eq}(T)}{T_g
s_c(T_g)} \right] \left[\frac{\gamma(T_g)}{\gamma(T)} \right]^2
\hspace{10mm} \nonumber \\ \hspace{10mm} = \left\{ \frac{T-T_K [1+
\ln(T/T_g)]}{T_g-T_K} \right\} \left[\frac{\gamma(T_g)}{\gamma(T)}
\right]^2, \hspace{5mm}
\label{nonArrh_rat}
\end{eqnarray}
where the second equality is found by computing $f_{eq}$ with the help
of the Angell's form of $\Delta c_p(T)$ (see the Appendix).  We plot
the ratio above in Fig.\ref{nonArr}. 
\begin{figure}[tb]
 \includegraphics[width=.85\columnwidth]{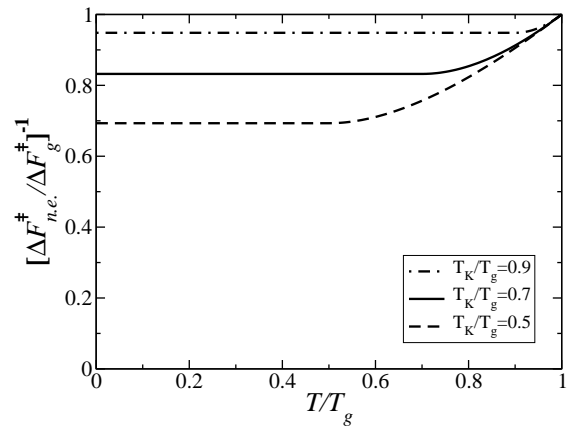}
 \caption{\label{nonArr} We plot the predicted variation of the
activation free energy versus inverse temperature below the glass
transition, as the ratio $(\Delta F_{n.e.}^\ddagger(T)/\Delta
F_g^\ddagger)^{-1}$ from Eq.(\ref{nonArrh_rat}) versus $T/T_g$. Only
below $T_K$ will the ratio be strictly constant, implying strictly
Arrhenius temperature dependence of the relaxation rate. The
temperature dependence of $\gamma$ is neglected. The figure is plotted
for a fragile material with $T_K/T_g$ (1 hr.) = 0.9, a less fragile
one with $T_K/T_g$= 0.7 and a strong substance with $T_K/T_g$= 0.5.  }
\end{figure}
We point out that although we have
highlighted the connection with more traditional treatments of aging
phenomenology, neither of the earlier approaches is exactly
commensurate with our theory. On the other hand we have shown the
expression for the nonequilibrium rate is very explicit once the
average energy of the sample is known. It therefore would not be
terribly difficult to use the full expression (\ref{k(x)g}) in a
dynamic treatment with the nonequilibrium energy as the indicator of
the ``fictive'' temperature. We do not carry out this analysis here
because it involves detailed numerical fitting for each system and the
quench history of each particular experiment.

\section{Nonexponentiality of Relaxation above and below $T_g$}

Within the RFOT theory the mosaic structure of the liquid gives rise
to dynamical heterogeneity and nonexponential relaxation. The driving
force for re-equilibration varies from mosaic cell to cell. This leads
to a range of activation barriers, $\delta F^\ddagger$. In the
canonical ensemble formulation, using the usual Landau formula, these
fluctuations depend on $\Delta c_p(T)$. Computing these fluctuations
allowed Xia and Wolynes to predict $\beta$ for a range of substances
\cite{XWbeta}. The same result can be obtained in the microcanonical
formulation, however here the origin is even more transparent: there
is a range of energies of the initial configurations for each mosaic
cell. Since the energy fluctuations also scale with $\Delta c_p(T)$,
above $T_g$ one obtains the same results as Xia and Wolynes previously
derived.

We first discuss how $\beta$ varies with temperature below $T_g$ in a
somewhat simplified approximation that makes the issues clear. We will
assume the energy fluctuations are small and the resulting barrier
distribution is Gaussian. We will see later that the results of this
analysis bound the magnitude of the {\em change} expected in $\beta$,
even when a more accurate barrier distribution is used.

Let us re-write Eq.(\ref{k(x)g}) for $\Delta F^\ddagger$ now including
a fluctuation for the initial energy per particle
\begin{equation} 
\Delta F_{n.e.}^\ddagger = \frac{\gamma^2}{4 [\phi_{in} -
f_{eq}+\delta \phi_{in}]}.
\label{dDFne}
\end{equation}
If $\delta \phi_{in}$ is small, we find the fluctuations in the
barrier height
\begin{equation} 
\delta \Delta F_{n.e.}^\ddagger = \frac{\gamma^2 (-\delta
\phi_{in})}{4 [\phi_{in} - f_{eq}]^2} = \frac{\Delta F_{n.e.}^\ddagger
(-\delta \phi_{in})}{\phi_{in} - \phi_{eq}+ T s_c(T)}.
\end{equation}
Since the structure is frozen at $T_g$ the typical fluctuation in
$\phi$ is the same as at $T_g$ and is determined by $\Delta c_p$ at
$T_g$. We see the ratio of the size of the $\Delta F^\ddagger$
fluctuations in the frozen, cooled state $\delta \Delta
F_{n.e.}^\ddagger$ to those found at $T_g$, $\delta \Delta
F_g^\ddagger$ is
\begin{equation} 
\frac{\delta \Delta F_{n.e.}^\ddagger}{\delta \Delta F_g^\ddagger} =
\frac{\Delta F_{n.e.}^\ddagger}{\Delta F_g^\ddagger} \frac{T_g
s_c(T_g)}{\phi_{in} - \phi_{eq}+ T s_c(T)}.
\label{dFdFratio}
\end{equation}
Both factors in this expression increase rather slowly below $T_g$,
and saturate at $T_K$.

\begin{figure}[tb]
 \includegraphics[width=.9\columnwidth]{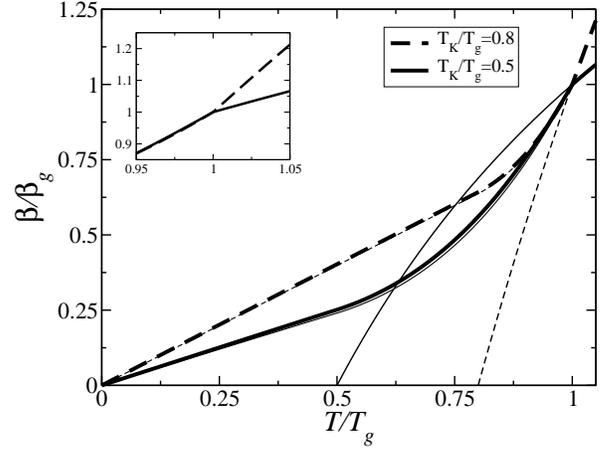}
 \caption{\label{b_T} The variation with temperature of the
nonequilibrium $\beta$ in comparison with $\beta(T_g)$ is shown as a
dashed line for a substance with $T_K/T_g$ (1 hr.) = 0.8,
characteristic of a fragile system and a solid line for a strong
system with $T_K/T_g$=.5. The two other lines indicates how the
equilibrium would vary with $T$. The approximte estimates from
Eq.(\ref{beta_app}) which exaggerate the variation are plotted as the
very thin lines, which nearly coincide with the more accurate
exression Eq.(\ref{beta}). The inset shows a magnified view of the
region near $T_g$, where only factual $\beta$'s, depicted by thick
lines in the main graph, are given.}
\end{figure}

How do these fluctuations translate into stretching exponents? Roughly
speaking, the relaxation function for a Gaussian distribution of
barriers is approximated by a stretched exponential with a value given
by \cite{XWbeta,Castaing}:
\begin{equation} 
\beta \simeq \left[1+ (\delta \Delta F^\ddagger/k_B T) \right]^{-1/2}.
\label{beta}
\end{equation}
If the liquid is ``strong'', $\Delta c_p$ is small so there are small
energy fluctuations leading to small $\delta \Delta F^\ddagger$. Thus
for strongly liquids $\beta$ remains near 1 until rather low
temperatures. If the fluctuations in $\phi$ are large (as they are for
very fragile systems) we would find instead
\begin{equation} 
\beta \simeq \frac{k_B T}{\delta \Delta F^\ddagger}.
\label{beta_app}
\end{equation}
This formula should thus give an overestimate for the variation of
$\beta$ with temperature. Using this estimate along with
Eqs.(\ref{dDFne}) and (\ref{dFdFratio}), we find
\begin{equation} 
\frac{\beta_{n.e.}(T)}{\beta(T_g)} = \frac{T}{T_g} \left[ \frac{\Delta
F_{n.e.}^\ddagger(T)}{\Delta F_g^\ddagger} \right]^{-2}
\left[\frac{\gamma(T)}{\gamma(T_g)} \right]^2.
\label{beta_ratio}
\end{equation}
A plot of this expression is shown in Fig \ref{b_T} (note the $\Delta
F_{n.e.}^\ddagger/\Delta F_g^\ddagger$ ratio has already been computed
in Eq.(\ref{nonArrh_rat})). 
At $T_K$ and below the term in parenthesis saturates and so, we find a
simple expression
\begin{equation} 
\frac{\beta_{n.e.}(T)}{\beta(T_g)} = \frac{T}{T_g} \left[ \frac{\Delta
F_{n.e.}^\ddagger(T_K)}{\Delta F_g^\ddagger} \right]^{-2}
\left[\frac{\gamma(T_K)}{\gamma(T_g)} \right]^2.
\label{beta_ratioK}
\end{equation}
Thus we see in general $\beta$ should fall as we cool but the effect
remains modest in the range where $T$ is greater than $T_K$. This
modestly cooled range is where the most detailed aging studies have
been reported \cite{Alegria,Leheny}. Although $\beta$ falls, indeed,
the rate of fall slows from its $T$ dependence above $T_g$ for fragile
substances, since according to the simple RFOT approximation of Xia
and Wolynes the equilibrium $\beta$ would vanish at $T_K$, while this
(approximate) nonequilibrium vanishes only at absolute zero. In
contrast, $\beta$ for strong substances, that was near unity and
$T$-independent above $T_g$, is predicted to show more pronounced
tempearature dependence below the glass transition (see the inset of
Fig.\ref{b_T}).  Thus we see the form of the relaxation is not the
same as the equilibrium relaxation at any ``fictive'' temperature. For
fragile liquids $T_K/T_g \simeq .8$, so if $\gamma$ is temperature
independent, a 40\% reduction is expected until $T_K$ is
reached. Often relaxation data in the aging regime have been fit with
the approximation $\beta = \beta(T_g)$. Alegria et al. suggest $\beta$
remains constant below $T_g$. The RFOT indicates this is a reasonable
zeroth order approximation. Alegria et al. have measured $\beta$ in
the regime $T_g >T> T_K$. These are difficult measurements and there
is scatter in the data.

We must bear in mind that the theoretical $\beta$ assumes that
measurements can scan over the complete relaxation process (even below
$T_g$!) but part of the relaxation is missed in experiment. Since the
glass only partly relaxes, in most experiments $\beta$ will appear to
be {\em closer} to one than if the full relaxation could be
followed. This effect of missing part of the distribution gives a
positive increment to $\beta$ at the glass transition, as pointed out
by Alegria et al. Such an increment on cooling is found.

After this artifactual increment in $\beta$, Alegria et al. actually
do find $\beta$ to slightly decrease upon cooling.  We must emphasize
that the arguments leading to Eq.(\ref{beta_ratio}) generally give an
overestimate of the variation of with temperature in the glass
state. First and most simply the approximation made in going from
Eq.(\ref{beta}) to Eq.(\ref{beta_app}) causes an overestimate. Second,
but more important, as discussed by Xia and Wolynes \cite{XWbeta} the
zeroth order Gaussian approximation for the barrier distribution is
not quantitative because the simple argument leading to that result
assumes the environment of a reconfiguring domain is temporally
fixed. Clearly if the domains surrounding a region that may
re-configure have themselves already changed before the
reconfiguration event, the library construction's premise of having a
fixed environment to the mosaic cell fails.  This change of
environment effect might be called a ``facilitation''
\cite{Andersen}. In any case this effect means the barrier height
distribution will be cut off on the high barrier side. A simple cutoff
distribution follows from the idea that domains slower than the most
probable rate would actually re-configure when their environmental
neighbors have changed; thus they actually will re-configure at nearly
the most probable rate, which has already been predicted by the RFOT
theory.  The resulting cutoff distribution for activation barrier:
\begin{equation}
P(\Delta F^\ddagger) = \left\{ \begin{array}{l} P_f(\Delta
F^\ddagger), \mbox{ for } \Delta F^\ddagger < \Delta F_0^\ddagger \\ C
\delta(\Delta F^\ddagger - \Delta F_0^\ddagger)
\end{array} \right. 
\label{Pf}
\end{equation}
- has been shown to reproduce the variation of the equilibrium at
$T_g$ with composition quite well \cite{XWbeta}. It also reproduces
the temperature dependence of $\beta$ in the equilibrated supercooled
liquid. The weight $C$ ensures the normalization of the
distribution. Clearly the cutoff again acts to dampen the variation of
with temperature, below $T_g$. Explicit calculations using
Eqs.(\ref{Pf}) and (\ref{dFdFratio}) for the fluctuation can be used
to study the detailed $T$ variation of $\beta$ in the nonequilibrium
glass, which may be relevant for deep quenches.


\section{Relaxation in Quenches below $T_K$}


We have emphasized dynamics in the glassy regime just below $T_g$ and
ranging down to $T_K$. In this regime dynamics is fast enough so that
significant relaxation is still accessible to detailed
experiments. The results we have obtained should hold to considerably
lower temperatures at least before much aging has actually occured.
Essentially the average rate will become Arrhenius below $T_K$ while
the breadth of relaxation times will continue to increase as $T$
approaches the absolute zero.  Quantitatively several effects may
intervene that are worthy of further study, however.  These effects
are 1) secondary relaxations, 2) fluctuations in mismatch energies and
changes in the ``wetting'' mechanism, and 3) quantum effects. We
comment on these in turn.

1) As a practical matter, less and less of the range of relaxation
times can be accessed on deep cooling during a typical laboratory
experiment - at the lowest temperatures only the fast end of the
relaxation time distribution can be accessed.  This part of the
distribution is caused by regions of high energy that correspond to a
small domain size. At the shorter length scale several effects that
are system specific, will occur.

This region of the relaxation time spectrum is often called the
$\beta$-relaxation. The term $\beta$-relaxation tends to bring to mind
universal characteristics and indeed the RFOT theory does for example
suggests a scale for the maximum rate corresponding to the cost of
overturning a single molecular unit; essentially this is
$\gamma_0$. On the other hand most glass formers have internal
structure and these inner parts or side-chains of the molecule will
have multiple conformations that can relax, in a specific ways as well
those predictable from RFOT alone \cite{slaving}. These ``rapid''
relaxations, also will slow with temperature, and might come into the
measurement window. Dyre has argued that just such a ``contamination''
of the main $\alpha$ aging process by a $\beta$ relaxation \cite{Dyre}
may explain some of the Leheny-Nagel measurements on glycerol
\cite{Leheny}.

2) Well below $T_K$ the mismatch energies may change their scaling
with size. The ``wetting'' mechanism relies on there being a
multiplicity of states to interpolate through the interface. Such a
multiplicity exists at $T_K$ but becomes less important with
decreasing $T$. This leads to a ``hardening'' of the interface
(increase in $\gamma$) or perhaps a crossover from $N^{1/2}$ scaling
to $N^{2/3}$ scaling of the mismatch energy. At the same time, rare
fluctuations of the barriers may allow some regions to relax more
rapidly than expected. Plotkin and Wolynes have analyzed just this
sort of effect in the context of ``buffing'' of energy landscapes in
protein \cite{buffing} and Lubchenko and Wolynes have studied a
similar effect in the quantum regime which allows a tail of resonant
tunneling states to appear \cite{LW}. Also well below $T_K$ - small
regions may reconfigure by crystallizing - an effect we ignore in the
present paper.

3) Eventually below $T_K$ classical barrier crossing will be
supplanted by quantum tunneling. This gives rise eventually to two
level systems \cite{LW} and the Boson peak \cite{LW_BP} which we have
discussed in detail elsewhere.

\vspace{-5mm}

\section{Aging and History Dependence}

The local energy landscape theory put forward in this paper predicts
the relaxation of a system where the statistics of the energies in the
local energy landscape libraries of the sample is assumed to be known
at any instant. The explicit formula for the typical relaxation rate
depends on the mean bulk energy of the cooperative regions and the
explicit formula for the stretching exponent $\beta$ also contains the
variance of these energies. In the quenched sample we have assumed
these statistics are characteristic of an equilibrium system at the
temperature $T_g$ where the system ``fell out of equilibrium''.  Since
the cooperative regions are large it is natural to assume these
statistics are Gaussian as we have done. In simple quench histories
$T_g$ can be estimated by the temperature where the apparent heat
capacity most rapidly falls during cooling. But more complicated
thermal histories are possible and even in a simple quench $T_g$
really must be determined self-consistently by the dynamics of the
system, itself. In general the statistics of the local landscape
libraries for a nonequilibrium system will be determined by the
system's dynamics and its detailed past thermal history.

\begin{figure}[tb]
 \includegraphics[width=.55\columnwidth]{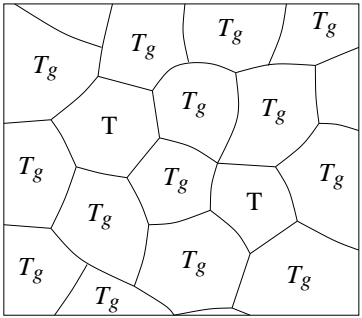}
 \caption{\label{inhomT} After a considerable period of aging well
below $T_g$ a patchwork of equilibrated and nonequilibrated mosaic
cells will be found. If the equilibrium energy at $T$ is more than a
standard deviation of the configurational energies at $T_g$, the
distribution of energies will be noticeably bimodal and the idea of a
single fictive temperature will break down. The unimodal distribution
with a single fictive temperature should be quite safe if $\Delta T =
T_g -T < \sqrt{k_B T^2/\Delta c_p N^*} = \delta T^*$. For $T_g$
relevant to 1 hr. quenches this gives $\delta T^*/T_g \simeq
0.07$. Most of the Alegria et al. \cite{Alegria} data lie in this
modest quenching range, while ``hyperquenched'' samples (with $\Delta
T \gg \delta T^*$) will often fall outside the allowed range of using
a single fictive temperature. When a sample has a two peaked
distribution of local energies, an ultra-slow component of relaxation
will arise. Notice that an equilibrated region at the temperature
$T=T_g - \delta T^*$ will relax on the tens to hundreds of hours scale
(using the relation that $\Delta E_{n.e.L.T.}^\ddagger \simeq \Delta
F_g^\ddagger$), if $\tau_g$ is taken to be one hour.  Percolation of
the ultra slow regions will lead to the possibility of observing
multiple length scales.  }
\end{figure}

The most general description of the nonequilibrium statistics is quite
complex since the bulk energies of any region can be considered
functions over the shape and size of domains. These functions might be
described as a set of fluctuating bulk energy fields, but the
resulting construction is complex. When the system is cooling, high
energy regions of size $N^*$ will be replaced by regions equilibrated
to the ambient temperature $T$.  Thus there will be a patchwork of
equilibrated and nonequilibrated mosaic cells (see Figure
\ref{inhomT}). 
The nature of the new statistics, after some transitions occur and
substantial aging has progressed, depends on how big is the difference
between the typical current energy of a region and the target
equilibrium energy. If the gap is big, a two peaked structure in the
distribution of local bulk energies will develop and the statistics
will be far from Gaussian: some regions that are newly equilibrated
will relax further at a (slower) rate characteristic of equilibrium at
$T$, while the other, not yet transformed regions will still relax at
the faster nonequilibrium rate discussed already. Such a situation, if
it arises, might account for ``ultra-slow'' relaxations which have
occasionally been reported in aging studies \cite{MillerMcPhail} (see
the Figure \ref{inhomT} caption for a criterion for such
behavior). The statistics in this case of very deep quenches studied
for very long times is very complex kinematically. Two peaked
distributions of local energies may also arise if the system is
abruptly but briefly heated from a low temperature state to a much
higher one. Fortunately in the more usual situation of modest
monotonic quenches we can expect the distribution of local energies to
remain unimodal.  In such cases, to first order then the distribution
will be characterized by a mean bulk energy per particle
$\bar{\epsilon}$! If the ambient temperature were to remain fixed,
$\bar{\epsilon}$ should relax to $\epsilon_{eq}(T)$ since at that
point detailed balance applied to the microscopic rates will replenish
any states from which the system locally escapes. (We remind the
reader that $\epsilon$ denotes a (library) free energy $\phi$ without
the vibrational entropy contribution.) The typical escape rate at any
value of $\bar{\epsilon}$ will be given by Eq.(\ref{k(x)g}) at ambient
temperature $T$. Thus the equation of motion for will $\bar{\epsilon}$
have the form
\begin{widetext}
\begin{equation}
\bar{\epsilon}-\epsilon_{eq}(T) = - \int_{-\infty}^t d t_1
\dot{\rho}(t-t_1; T(t), \tau[\bar{\epsilon}(t_1), T(t_1)]) \:
\{\bar{\epsilon}(t_1) - \epsilon_{eq}[T(t_1)]\}.
\end{equation}
\end{widetext}
where $\rho$ is the relaxation function when the statistics of
libraries are known and fixed by $\bar{\epsilon}(t_1)$, $T(t_1)$ is
the ambient temperature at time $t_1$ - we assume vibrational energies
equilibrate quickly by thermal conduction. $\rho$ not only depends on
$T(t_1)$ and $\bar{\epsilon}$ but contains the parameter $\beta$ which
in turn depends on the variance of energies of the local region. As a
first approximation the variance can be taken as that characteristic
of an equilibrium system at a temperature $T^*$ such that
\begin{equation}
\epsilon_{eq}(T^*) = \bar{\epsilon}.
\end{equation}
In this approximation, we may use Eq.(\ref{beta_ratio}) for
\begin{equation}
\beta \equiv \beta_{in}[T(t),T(\bar{\epsilon})].
\end{equation}
We see that if the shape of the distribution of local libraries does
not change much, the RFOT picture leads to a situation where a single
parameter $\bar{\epsilon}$ suffices to characterize the
system. $T^*(\bar{\epsilon})$ thus essentially fixes the ``fictive
temperature'' $T_f$ in the NTM phenomenology, albeit with different
expressions for $\tau(T, T_f)$ and $\beta=\beta(T, T_f)$.  The
non-Gaussian statistics alluded to earlier, however suggests the use
of a single fictive temperature is only approximate and that a more
complete characterization of the statistics may be needed. Multiple
fictive temperatures defining the higher moments could in principle be
defined. In addition spatial fluctuations of fictive temperatures are
needed to capture the co-existence of equilibrated and nonequilibrated
domains in the mosaic structure. At least for moderate quenches it may
be possible to ignore the spatial inhomogenity and merely monitor the
fluctuations in energies of domains as a secondary variable.

\vspace{-5mm}

\section{Summary}

We described a local energy landscape theory of the dynamics of
supercooled liquids and glasses. In the equilibrated supercooled
regime this theory is just a microcanonical ensemble reformulation of
the random first order transition theory and its notion of entropic
droplets. New results are obtained in the aging regime of
nonequilibrium quenched glasses. The key equation is (\ref{k(x)g})
which shows it is the difference between the equilibrated free energy
at the quench temperature and the initial free energy of the
particular frozen state that drives motions. The theory approximately
reproduces the phenomenology of Narayanaswamy and Moynihan and
Tool. Thus the nonlinearity parameter in NMT theory can be
calculated. This parameter is shown to be correlated with the
supercooled liquid's fragility in agreement with experiment. This
correlation is quantitatively very similar to that obtained by the
Hodge-Scherer-Adam-Gibbs extrapolation that assumes configurational
entropy is fixed at $T_g$. Deviations from that extrapolation which
assumes Arrhenius behavior in the glassy state are predicted
however. The previously puzzling, modest non-Arrhenius temperature
dependence of relaxation observed within the glassy state is explained
by the RFOT theory, although it is a small effect. The variations of
nonexponentiality of relaxation in the glassy state are predicted but
are also rather small in the moderately quenched regime. The
comparison of nonexponentiality with experiment is less conclusive
than the comparison of mean relaxation rates however, owing to the
difficulty of accessing the complete relaxation behavior during the
quench.

One advantage of the aging theory based on RFOT theory is that in
principle the behavior upon very deep quenches is predicted. Most
importantly $\beta$ is predicted to continue to decrease with quench
temperature. We hope that more experiments in this regime will be
done. We have noted however that some non-universal effects may enter
for such quenches. Also the kinematics of deep quenches may be complex
owing to spatial fluctuations of fictive temperature predicted by our
theory.

Although the present approach justifies to some extent the use of a
single fictive temperature to characterize the glassy state the
limitations of this idea have been made clear. A straightforward
extension of the usual formulation to include fluctuations in the
moderately cooled regime more fully was proposed. In the strongly
cooled regime, the theory predicts patches of especially slowly
relaxing regions will appear. This prediction may be tested by single
molecule imaging approaches \cite{RussellIsraeloff,VandenBout}.

Acknowledgement: Peter G. Wolynes would like to thank Xiaoyu Xia for
early discussions on this topic. The work was supported by the
National Science Foundation grant CHE0317017. Vassiliy Lubchenko is
grateful for the kind support and indulgence of Robert Silbey at
M.I.T.

\vspace{-5mm}

\appendix

\section*{Appendix}

\setcounter{equation}{0}
\renewcommand{\theequation}{A.\arabic{equation}}

Several auxiliary results are derived in this Appendix (here,
$k_B=1$). First, we derive formula (\ref{nonArrh_rat}) from the main
text. Using $s = -\partial f/\partial T$, one has (neglecting
differences in vibrational entropy)
\begin{eqnarray}
f_{eq}(T) &=& \phi_K - \int_{T_K}^{T} s dT \nonumber \\ &=& \phi_K -
  \Delta c_p(T_g) T_g \left(\frac{T-T_K}{T_K} - \ln \frac{T}{T_K}
  \right), \hspace{7mm}
\label{-10} 
\end{eqnarray}
where we have used Angell's empirical form \cite{XW} for the
configurational entropy, also used in the main text: $s_c(T) = \Delta
c_p(T_g) T_g(1/T_K-1/T), \mbox{ } \Delta c_p(T) = \Delta c_p(T_g)
(T_g/T)$. Noting that $f_{eq}(T_g) = \phi_g - T_g s_c(T_g)$ fixes the
ideal glass state energy $\phi_K$:
\begin{equation}
\phi_K = \phi_g - \Delta c_p(T_g) T_g \ln(T_g/T_K).
\label{-9}
\end{equation}
Eqs.(\ref{-10}) and (\ref{-9}) immediately yield
Eq.(\ref{nonArrh_rat}).

Next, we derive the more accurate expression for the
$\beta_{n.e.}(T)/\beta(T_g)$, as follows from Eq.(\ref{beta}). We also
provide the equilibrium value $\beta_{eq}$ of the non-exponentiality
parameter $\beta$, previously obtained by Xia and Wolynes \cite{XW},
in terms of experimental parameters $T_K/T_g$ and
$\ln(\tau_Q/\tau_{micro})$, used throughout this paper.

First we derive $\beta_{eq}$. According to Ref.\cite{XW}, the barrier
fluctuations are directly related to the local fluctuations in the
configurationlal entropy at the scale $N^*$ corresponding to a local
equilibrium unit, of which the global composite landscape is
comprised: $\delta \Delta F^\ddagger/\Delta F^\ddagger = \delta
S_c/S_c = \sqrt{\Delta c_P N^*}/s_c N^*$. This yields:
\begin{equation}
\frac{\delta \Delta F^\ddagger}{T} = \frac{\Delta F^\ddagger}{T}
\frac{\sqrt{\Delta c_P}}{s_c \sqrt{N^*}}.
\end{equation}

It directly follows from $F^\ddagger(N) = \gamma \sqrt{N} - Ts_cN$,
$(dF/dN)_{N=N^\ddagger} = 0$, and $F(N^*) = 0$ that
\begin{equation}
\frac{\Delta F^\ddagger}{T} = \frac{(\gamma/T)^2}{4s_c}
\label{1}
\end{equation}
and
\begin{equation}
\sqrt{N^*} = \frac{\gamma/T}{s_c}.
\end{equation}
Using Eq.(\ref{tQ}), Eq.(\ref{1}) and the temperature independence of
the $\gamma/T$ ratio in equilibrium \cite{XW}, one gets
\begin{equation}
(\gamma/T) = (\gamma/T)_{T=T_g} = 2 \sqrt{s_c(T_g)
\ln(\tau_Q/\tau_{micro})}.
\end{equation}
With the help of the equations above and Angell's empirical form for
the configurational entropy one easily obtains that
\begin{equation}
\frac{\delta \Delta F^\ddagger}{T} =
\frac{\sqrt{\ln(\tau_Q/\tau_{micro})}}{2}
\frac{\sqrt{1/T_K-1/T_g}}{\sqrt{T}(1/T_K-1/T)}.
\label{2}
\end{equation}
This and Eq.(\ref{beta}) can be used to compute the
$\beta_{eq}(T)/\beta(T_g)$ ratio.

The non-equilibrium $\beta$ requires even less effort. Using
Eq.(\ref{dFdFratio}), one gets
\begin{equation}
\frac{\delta \Delta F^\ddagger_{n.e.}}{\delta \Delta F^\ddagger_g} =
\left[ \frac{F_{n.e.}^\ddagger(T)}{F_g^\ddagger} \right]^2 \left[
\frac{\gamma(T_g)}{\gamma(T)} \right]^2.
\end{equation}
Hence,
\begin{equation}
\beta_{n.e.}(T) = \left[ 1+ \left(\frac{\delta \Delta
F^\ddagger_g}{T}\right)^2
\left[\frac{F_{n.e.}^\ddagger(T)}{F_g^\ddagger} \right]^4 \left[
\frac{\gamma(T_g)}{\gamma(T)} \right]^4 \right]^{-1/2}.
\end{equation}
The ratio $F_{n.e.}^\ddagger(T)/F_g^\ddagger$ was computed in the
beginning of the Appendix and given in the main text as
Eq.(\ref{nonArrh_rat}); $\delta \Delta F^\ddagger(T_g)$ is obtained
from Eq.(\ref{2}) at $T=T_g$.


 \end{document}